
\catcode`@=11

\def\oneandahalfspace{\baselineskip=1.15\normalbaselineskip plus 1pt
\lineskip=2pt\lineskiplimit=1pt}

\def\nl{\hfil\break}

\def\nofirstpagenoten{\nopagenumbers\footline={\ifnum\pageno>1\tenrm
\hss\folio\hss\fi}}
\def\nofirstpagenotwelve{\nopagenumbers\footline={\ifnum\pageno>1\twelverm
\hss\folio\hss\fi}}
\def\leaderfill{\leaders\hbox to 1em{\hss.\hss}\hfill}
\def\ft#1#2{{\textstyle{{#1}\over{#2}}}}
\def\frac#1/#2{\leavevmode\kern.1em
\raise.5ex\hbox{\the\scriptfont0 #1}\kern-.1em/\kern-.15em
\lower.25ex\hbox{\the\scriptfont0 #2}}
\def\sfrac#1/#2{\leavevmode\kern.1em
\raise.5ex\hbox{\the\scriptscriptfont0 #1}\kern-.1em/\kern-.15em
\lower.25ex\hbox{\the\scriptscriptfont0 #2}}

\parindent=20pt
\def\narrow{\advance\leftskip by 40pt \advance\rightskip by 40pt}

\def\AB{\bigskip
        \centerline{\bf ABSTRACT}\medskip\narrow}
\def\nonarrower{\advance\leftskip by -40pt\advance\rightskip by -40pt}
\def\AE{\bigskip\nonarrower}

\def\boxit#1{\vbox{\hrule\hbox{\vrule\kern3pt
        \vbox{\kern3pt#1\kern3pt}\kern3pt\vrule}\hrule}}

\def\gtorder{\mathrel{\raise.3ex\hbox{$>$}\mkern-14mu
             \lower0.6ex\hbox{$\sim$}}}
\def\ltorder{\mathrel{\raise.3ex\hbox{$<$}|mkern-14mu
             \lower0.6ex\hbox{\sim$}}}
\def\dalemb#1#2{{\vbox{\hrule height .#2pt
        \hbox{\vrule width.#2pt height#1pt \kern#1pt
                \vrule width.#2pt}
        \hrule height.#2pt}}}
\def\square{\mathord{\dalemb{4.9}{5}\hbox{\hskip1pt}}}

\font\fourteentt=cmtt10 scaled \magstep2
\font\fourteenbf=cmbx12 scaled \magstep1
\font\fourteenrm=cmr12 scaled \magstep1
\font\fourteeni=cmmi12 scaled \magstep1
\font\fourteenssr=cmss12 scaled \magstep1
\font\fourteenmbi=cmmib10 scaled \magstep2
\font\fourteensy=cmsy10 scaled \magstep2
\font\fourteensl=cmsl12 scaled \magstep1
\font\fourteenex=cmex10 scaled \magstep2
\font\fourteenit=cmti12 scaled \magstep1
\font\twelvett=cmtt12 \font\twelvebf=cmbx12
\font\twelverm=cmr12  \font\twelvei=cmmi12
\font\twelvessr=cmss12 \font\twelvembi=cmmib10 scaled \magstep1
\font\twelvesy=cmsy10 scaled \magstep1
\font\twelvesl=cmsl12 \font\twelveex=cmex10 scaled \magstep1
\font\twelveit=cmti12
\font\tenssr=cmss10 \font\tenmbi=cmmib10
 
 \font\ninebf=cmbx9
\font\ninerm=cmr9  \font\ninei=cmmi9
\font\ninesy=cmsy9 \font\ninessr=cmss9
\font\ninembi=cmmib10 scaled 900
\font\eightit=cmti8 \font\eightsl=cmsl8
\font\eighttt=cmtt8 \font\eightbf=cmbx8
\font\eightrm=cmr8  \font\eighti=cmmi8
\font\eightsy=cmsy8 \font\eightex=cmex10 scaled 800
\font\eightssr=cmss8 \font\eightmbi=cmmib10 scaled 800
 
\font\sevenbf=cmbx7 \font\sevenrm=cmr7 \font\seveni=cmmi7
\font\sevensy=cmsy7 
\font\sevenssr=cmss9 scaled 778 \font\sevenmbi=cmmib10 scaled 700
 
 \font\sixbf=cmbx7 scaled 875
\font\sixrm=cmr6  \font\sixi=cmmi6
\font\sixsy=cmsy6 \font\sixssr=cmss8 scaled 750
\font\sixmbi=cmmib10 scaled 600
\font\fivessr=cmss8 scaled 625  \font\fivembi=cmmib10 scaled 500

\newskip\ttglue
\newfam\ssrfam
\newfam\mbifam

\mathchardef\alpha="710B
\mathchardef\beta="710C
\mathchardef\gamma="710D
\mathchardef\delta="710E
\mathchardef\epsilon="710F
\mathchardef\zeta="7110
\mathchardef\eta="7111
\mathchardef\theta="7112
\mathchardef\iota="7113
\mathchardef\kappa="7114
\mathchardef\lambda="7115
\mathchardef\mu="7116
\mathchardef\nu="7117
\mathchardef\xi="7118
\mathchardef\pi="7119
\mathchardef\rho="711A
\mathchardef\sigma="711B
\mathchardef\tau="711C
\mathchardef\upsilon="711D
\mathchardef\phi="711E
\mathchardef\chi="711F
\mathchardef\psi="7120
\mathchardef\omega="7121
\mathchardef\varepsilon="7122
\mathchardef\vartheta="7123
\mathchardef\varpi="7124
\mathchardef\varrho="7125
\mathchardef\varsigma="7126
\mathchardef\varphi="7127
\mathchardef\partial="7140

\def\fourteenpoint{\def\rm{\fam0\fourteenrm}
\textfont0=\fourteenrm \scriptfont0=\tenrm \scriptscriptfont0=\sevenrm
\textfont1=\fourteeni \scriptfont1=\teni \scriptscriptfont1=\seveni
\textfont2=\fourteensy \scriptfont2=\tensy \scriptscriptfont2=\sevensy
\textfont3=\fourteenex \scriptfont3=\fourteenex \scriptscriptfont3=\fourteenex
\def\it{\fam\itfam\fourteenit} \textfont\itfam=\fourteenit
\def\sl{\fam\slfam\fourteensl} \textfont\slfam=\fourteensl
\def\bf{\fam\bffam\fourteenbf} \textfont\bffam=\fourteenbf
\scriptfont\bffam=\tenbf \scriptscriptfont\bffam=\sevenbf
\def\tt{\fam\ttfam\fourteentt} \textfont\ttfam=\fourteentt
\def\ssr{\fam\ssrfam\fourteenssr} \textfont\ssrfam=\fourteenssr
\scriptfont\ssrfam=\tenmbi \scriptscriptfont\ssrfam=\sevenmbi
\def\mbi{\fam\mbifam\fourteenmbi} \textfont\mbifam=\fourteenmbi
\scriptfont\mbifam=\tenmbi \scriptscriptfont\mbifam=\sevenmbi
\tt \ttglue=.5em plus .25em minus .15em
\normalbaselineskip=16pt
\bigskipamount=16pt plus5pt minus5pt
\medskipamount=8pt plus3pt minus3pt
\smallskipamount=4pt plus1pt minus1pt
\abovedisplayskip=16pt plus 4pt minus 12pt
\belowdisplayskip=16pt plus 4pt minus 12pt
\abovedisplayshortskip=0pt plus 4pt
\belowdisplayshortskip=9pt plus 4pt minus 6pt
\parskip=5pt plus 1.5pt
\twelvefoot
\setbox\strutbox=\hbox{\vrule height12pt depth5pt width0pt}
\let\sc=\tenrm
\let\big=\fourteenbig \normalbaselines\rm}
\def\fourteenbig#1{{\hbox{$\left#1\vbox to12pt{}\right.\n@space$}}
\def\square{\mathord{\dalemb{6.8}{7}\hbox{\hskip1pt}}}}

\def\twelvepoint{\def\rm{\fam0\twelverm}
\textfont0=\twelverm \scriptfont0=\ninerm \scriptscriptfont0=\sevenrm
\textfont1=\twelvei \scriptfont1=\ninei \scriptscriptfont1=\seveni
\textfont2=\twelvesy \scriptfont2=\ninesy \scriptscriptfont2=\sevensy
\textfont3=\twelveex \scriptfont3=\twelveex \scriptscriptfont3=\twelveex
\def\it{\fam\itfam\twelveit} \textfont\itfam=\twelveit
\def\sl{\fam\slfam\twelvesl} \textfont\slfam=\twelvesl
\def\bf{\fam\bffam\twelvebf} \textfont\bffam=\twelvebf
\scriptfont\bffam=\ninebf \scriptscriptfont\bffam=\sevenbf
\def\tt{\fam\ttfam\twelvett} \textfont\ttfam=\twelvett
\def\ssr{\fam\ssrfam\twelvessr} \textfont\ssrfam=\twelvessr
\scriptfont\ssrfam=\ninessr \scriptscriptfont\ssrfam=\sevenssr
\def\mbi{\fam\mbifam\twelvembi} \textfont\mbifam=\twelvembi
\scriptfont\mbifam=\ninembi \scriptscriptfont\mbifam=\sevenmbi
\tt \ttglue=.5em plus .25em minus .15em
\normalbaselineskip=14pt
\bigskipamount=14pt plus4pt minus4pt
\medskipamount=7pt plus2pt minus2pt
\abovedisplayskip=14pt plus 3pt minus 10pt
\belowdisplayskip=14pt plus 3pt minus 10pt
\abovedisplayshortskip=0pt plus 3pt
\belowdisplayshortskip=8pt plus 3pt minus 5pt
\parskip=3pt plus 1.5pt
\tenfoot
\setbox\strutbox=\hbox{\vrule height10pt depth4pt width0pt}
\let\sc=\ninerm
\let\big=\twelvebig \normalbaselines\rm}
\def\twelvebig#1{{\hbox{$\left#1\vbox to10pt{}\right.\n@space$}}
\def\square{\mathord{\dalemb{5.9}{6}\hbox{\hskip1pt}}}}

\def\tenpoint{\def\rm{\fam0\tenrm}
\textfont0=\tenrm \scriptfont0=\sevenrm \scriptscriptfont0=\fiverm
\textfont1=\teni \scriptfont1=\seveni \scriptscriptfont1=\fivei
\textfont2=\tensy \scriptfont2=\sevensy \scriptscriptfont2=\fivesy
\textfont3=\tenex \scriptfont3=\tenex \scriptscriptfont3=\tenex
\def\it{\fam\itfam\tenit} \textfont\itfam=\tenit
\def\sl{\fam\slfam\tensl} \textfont\slfam=\tensl
\def\bf{\fam\bffam\tenbf} \textfont\bffam=\tenbf
\scriptfont\bffam=\sevenbf \scriptscriptfont\bffam=\fivebf
\def\tt{\fam\ttfam\tentt} \textfont\ttfam=\tentt
\def\ssr{\fam\ssrfam\tenssr} \textfont\ssrfam=\tenssr
\scriptfont\ssrfam=\sevenssr \scriptscriptfont\ssrfam=\fivessr
\def\mbi{\fam\mbifam\tenmbi} \textfont\mbifam=\tenmbi
\scriptfont\mbifam=\sevenmbi \scriptscriptfont\mbifam=\fivembi
\tt \ttglue=.5em plus .25em minus .15em
\normalbaselineskip=12pt
\bigskipamount=12pt plus4pt minus4pt
\medskipamount=6pt plus2pt minus2pt
\abovedisplayskip=12pt plus 3pt minus 9pt
\belowdisplayskip=12pt plus 3pt minus 9pt
\abovedisplayshortskip=0pt plus 3pt
\belowdisplayshortskip=7pt plus 3pt minus 4pt
\parskip=0.0pt plus 1.0pt
\eightfoot
\setbox\strutbox=\hbox{\vrule height8.5pt depth3.5pt width0pt}
\let\sc=\eightrm
\let\big=\tenbig \normalbaselines\rm}
\def\tenbig#1{{\hbox{$\left#1\vbox to8.5pt{}\right.\n@space$}}
\def\square{\mathord{\dalemb{4.9}{5}\hbox{\hskip1pt}}}}

\def\eightpoint{\def\rm{\fam0\eightrm}
\textfont0=\eightrm \scriptfont0=\sixrm \scriptscriptfont0=\fiverm
\textfont1=\eighti \scriptfont1=\sixi \scriptscriptfont1=\fivei
\textfont2=\eightsy \scriptfont2=\sixsy \scriptscriptfont2=\fivesy
\textfont3=\eightex \scriptfont3=\eightex \scriptscriptfont3=\eightex
\def\it{\fam\itfam\eightit} \textfont\itfam=\eightit
\def\sl{\fam\slfam\eightsl} \textfont\slfam=\eightsl
\def\bf{\fam\bffam\eightbf} \textfont\bffam=\eightbf
\scriptfont\bffam=\sixbf \scriptscriptfont\bffam=\fivebf
\def\tt{\fam\ttfam\eighttt} \textfont\ttfam=\eighttt
\def\ssr{\fam\ssrfam\eightssr} \textfont\ssrfam=\eightssr
\scriptfont\ssrfam=\sixssr \scriptscriptfont\ssrfam=\fivessr
\def\mbi{\fam\mbifam\eightmbi} \textfont\mbifam=\eightmbi
\scriptfont\mbifam=\sixmbi \scriptscriptfont\mbifam=\fivembi
\tt \ttglue=.5em plus .25em minus .15em
\normalbaselineskip=9pt
\bigskipamount=9pt plus3pt minus3pt
\medskipamount=5pt plus2pt minus2pt
\abovedisplayskip=9pt plus 3pt minus 9pt
\belowdisplayskip=9pt plus 3pt minus 9pt
\abovedisplayshortskip=0pt plus 3pt
\belowdisplayshortskip=5pt plus 3pt minus 4pt
\parskip=0.0pt plus 1.0pt
\setbox\strutbox=\hbox{\vrule height8.5pt depth3.5pt width0pt}
\let\sc=\sixrm
\let\big=\eightbig \normalbaselines\rm}
\def\eightbig#1{{\hbox{$\left#1\vbox to6.5pt{}\right.\n@space$}}
\def\square{\mathord{\dalemb{3.9}{4}\hbox{\hskip1pt}}}}

\def\vfootnote#1{\insert\footins\bgroup\footsuite
    \interlinepenalty=\interfootnotelinepenalty
    \splittopskip=\ht\strutbox
    \splitmaxdepth=\dp\strutbox \floatingpenalty=20000
    \leftskip=0pt \rightskip=0pt \spaceskip=0pt \xspaceskip=0pt
    \textindent{#1}\footstrut\futurelet\next\fo@t}
\def\hangfootnote#1{\edef\@sf{\spacefactor\the\spacefactor}#1\@sf
    \insert\footins\bgroup\footsuite
    \let\par=\endgraf
    \interlinepenalty=\interfootnotelinepenalty
    \splittopskip=\ht\strutbox
    \splitmaxdepth=\dp\strutbox \floatingpenalty=20000
    \leftskip=0pt \rightskip=0pt \spaceskip=0pt \xspaceskip=0pt
    \smallskip\item{#1}\bgroup\strut\aftergroup\@foot\let\next}
\def\footsuite{}
\def\twelvefoot{\def\footsuite{\twelvepoint}}
\def\tenfoot{\def\footsuite{\tenpoint}}
\def\eightfoot{\def\footsuite{\eightpoint}}
\catcode`@=12

\twelvepoint
\oneandahalfspace
\nofirstpagenotwelve
\hoffset=-.2in
\voffset=-.2in
\hsize=6.8in
\vsize=9.4in
\def\h{{1\over 2}}

\def\p{\partial}

\def\tr{{\rm tr}}
\def\bs{{\mbi\sigma}}

\line{\hfil UG-5/92}
\line{\hfil CTP-TAMU-63/92}
\line{\hfil DAMTP/R-92/32}
\line{\hfil Imperial/TP/91-92/30}
\line{\hfil SISSA 157/92/EP}
\line{\hfil hep-th/9212037}
\vskip 0.4truecm
\centerline{\bf $U(1)$--Extended Gauge Algebras in $p$-Loop Space}
\vskip 1truecm
\centerline{{\bf E. Bergshoeff$^1$, R. Percacci$^2$,
E. Sezgin$^3$}\footnote{$^*$}{\tenfoot\sl Work supported in
part by NSF grant PHY-9106593.},}
\centerline{{\bf K.S. Stelle$^4$}\footnote{$^{\star}$}{\tenfoot\sl Supported in
part by the Commission of the European Communities under Contract
SC1*-CT91-0674.}{\bf and P.K. Townsend$^5$}} \bigskip\bigskip
{\it
\leftline{$^1$ Institute for Theoretical Physics, Nijenborgh 4,
9747 Groningen, The Netherlands}
\leftline{$^2$ International School for Advanced Studies,
via Beirut 4, 34014 Trieste, Italy}
\leftline{$^3$ Center for Theoretical Physics, Texas A\&M University,
College Station, Texas 77843}
\leftline{$^4$ The Blackett Laboratory, Imperial College, London
SW7 2BZ, England}
\leftline{$^5$ DAMTP, University of Cambridge, Silver Street,
Cambridge, CB3 9EW, England}
}
\vskip 0.8truecm
\AB
We consider, for $p$ odd, a $p$--brane coupled to a $(p+1)$th rank background
antisymmetric tensor field and to background Yang-Mills (YM) fields {\it via} a
Wess-Zumino term. We obtain the generators of antisymmetric tensor and
Yang-Mills gauge transformations acting on $p$--brane wavefunctionals
(functions on `$p$-loop space'). The Yang-Mills generators do not form a closed
algebra by themselves; instead, the algebra of Yang-Mills and antisymmetric
tensor generators is a $U(1)$ extension of the usual algebra of Yang-Mills
gauge
transformations. We construct the $p$-brane's Hamiltonian and thereby find
gauge-covariant functional derivatives acting on $p$--brane wavefunctionals
that commute with the YM and $U(1)$ generators.
\AE

\vfill\eject
\noindent {\bf 1. Introduction }
\medskip

In many supergravity theories the graviton supermultiplet includes a
$(p+1)$-form gauge potential that couples naturally to a $p$-dimensional
extended object, {\it i.e.}\ a $p$--brane. Two examples of interest are a
string
coupled to ten-dimensional ($d=10$) supergravity in the 2-form formulation and
a
fivebrane coupled to $d=10$ supergravity in the 6-form formulation. A feature
of the two-form formulation is that the two-form potential acquires a
non-trivial Yang-Mills (YM) transformation when YM fields are included [1].
Although there is no analogous `anomalous' YM variation of the six-form in the
six-form formulation of {\sl classical} supergravity/YM theory, such a
variation is required for anomaly cancellation in the quantum theory [2]. In
both formulations, therefore, one finds that the YM algebra is modified in the
sense that a commutator of two YM transformations yields not only another YM
transformation but also an antisymmetric tensor gauge transformation. One would
expect that, upon quantization of the $p$--brane in
the YM and antisymmetric tensor background, this modified algebra should be
realized in terms of functional differential operators acting on the $p$--brane
wavefunctional. This is a function on $p$-loop space, {\it i.e.}\ the space of
maps of the $p$--brane to spacetime. Since the antisymmetric tensor
transformation of the $(p+1)$-form on spacetime is equivalent to a $U(1)$
transformation of an associated one-form on $p$-loop space [3], one expects the
modified algebra to be a $U(1)$ extension of the algebra of YM gauge
transformations.

To investigate this point one first needs an action for the $p$--brane
coupled to these background fields. For $p$ {\sl odd}, which includes the
$p=1$ and $p=5$ cases under discussion, such an action has been proposed
for the bosonic $p$--brane [4]. This action describes a $p$--brane
propagating in a curved background locally diffeomorphic to $M\times G$,
where $M$ is spacetime and $G$ is a group manifold, and includes a
Wess-Zumino (WZ) term. (The coupling to target spacetime
gauge fields considered here is analogous to, but should not be directly
confused with, the gauging of Wess-Zumino terms {\it via} world-volume
gauge fields as considered in [5].) For $p=1$ this action reduces to the
bosonic sector of an action for the heterotic string appearing in earlier
work [6]. More recently, a set of YM generators acting on string
wavefunctionals was deduced from this action and shown to satisfy an
affine Ka{\v c}-Moody algebra [7]. The central charge appearing in this
algebra can be interpreted as the eigenvalue of the $U(1)$ generator
associated with the antisymmetric tensor gauge transformation [8]. One
purpose of this paper is to obtain the analogous results for all odd p
(partial results have appeared [9] during the course of writing this
paper). Our method is also novel; from a path integral representation of
the $p$--brane wavefunctional $\Psi$, together with a careful treatment
of boundary terms, we find that $\Psi$ satisfies the conditions
$$
G_\epsilon\Psi=0\qquad G_\Lambda\Psi=0, \eqno (1.1)
$$
with specific operators $G_\epsilon$ and $G_\Lambda$. These operators are
not purely operators on p-loop space because they also contain functional
derivatives with respect to the background gauge fields; consequently,
eqs.\ (1.1) do not constrain $\Psi$ for given fixed background fields but
simply determine the response of the wavefunctional to a gauge
transformation (if the background fields were treated as dynamical
variables, eqs.\ (1.1) could be interpreted as continuity equations.). The
consistency of these equations follows from the fact that the operators
$G_\epsilon$ and $G_\Lambda$ form a {\sl closed} algebra, for which the
only non-zero commutator is $$ [G_{\epsilon_1}, G_{\epsilon_2}]=
G_{[\epsilon_1,\epsilon_2]} -G_\Lambda \eqno (1.2) $$
where
$$
\Lambda = k_p\omega^2_p(A,\epsilon_1,\epsilon_2)
\eqno (1.3)
$$
is a $p$--form 2-cocycle of the algebra of YM gauge transformations and
$k_p$ is a normalisation constant. For $p=1,3,5$, which are the principal
cases of interest (see section 3 for details of the notation),
$$
\eqalign{
\omega^2_1(A,\epsilon_1,\epsilon_2)=&\ -2\tr\,\epsilon_1 d\epsilon_2\cr
\omega^2_3(A,\epsilon_1,\epsilon_2)=&\ -\tr\,\{d\epsilon_1,d\epsilon_2\}A\cr
\omega^2_5(A,\epsilon_1,\epsilon_2)=&\ {1\over15}\tr\,
(5F-3A^2)\left[2A\{d\epsilon_1, d\epsilon_2\}-
d\epsilon_1 A d\epsilon_2+d\epsilon_2 A d\epsilon_1\right]\ .\cr}
\eqno (1.4)
$$
A special feature of the string is that in this case the cocycle (1.3) is
background-field independent, so the algebra defined by (1.2) is a Lie algebra.
For $p>1$ the structure `constants' of the algebra are background-field
dependent.

Note that these results are similar to those found from the analysis of chiral
anomalies in the Hamiltonian formalism of $(p+1)$-dimensional gauge
theories [10, 11], but there are several differences. One is that here
the gauge fields are not dynamical so the anomalies in question are of
`sigma-model' type. A more significant, but related, difference is the
presence of the $(p+1)^{\rm th}$ rank antisymmetric tensor and its
`anomalous' YM transformation (without which there would be no background
gauge invariance of the string action; {\it cf.}\ the sigma-model anomaly
cancellation in the fermionic formulation of the heterotic string [12]).
With the antisymmetric tensor field, the algebra is anomaly-free, despite
the central extension, in the sense that equations (1.1) do not imply a
vanishing wavefunctional.

Since our $p$-brane action is worldvolume reparametrization invariant, its
canonical Hamiltonian is a sum of constraints ({\it cf.}\ General Relativity).
The `Hamiltonian' constraint function associated with time reparametrizations
is a quadratic function of the momenta conjugate to the worldvolume fields
$x^\mu$ and $y^m$ (these being maps from the worldvolume to $M$ and $G$
respectively). The invariance of the Hamiltonian under background field
transformations is a consequence of the invariance of particular linear
combinations of the momenta that become, upon quantization, functional YM and
$U(1)$ `covariant' derivatives acting on $p$-brane wavefunctionals. There are
{\sl two} covariant derivatives, ${\cal D}_\mu$ and ${\cal D}_a \equiv L_a{}^m
{\cal D}_m$, corresponding to covariant differentiation with respect to $x^\mu$
or to $y^m$, respectively ($L_a{}^m$ are the components of the left-invariant
Killing vectors on $G$). The first of these was introduced for the string in
Ref.\ [7], where it was used to derive dynamical equations for the
background fields {\it via} the principle of lightlike integrability.
Existence arguments for ${\cal D}_\mu$ in the general case have also been
given [13]. From the point of view of this paper the covariant derivatives
are functional differential operators on $p$--loop space satisfying
$$
\eqalign{
[{\cal D}_\mu, G_\Lambda]=0 \qquad &[{\cal D}_\mu, G_\epsilon]=0\cr
[{\cal D}_a, G_\Lambda]=0 \qquad &[{\cal D}_a, G_\epsilon]= \epsilon^b
 f^d{}_{ab}{\cal
D}_d\ .\cr}
\eqno (1.5)
$$
Another result of this paper is the construction of these covariant derivatives
for all odd $p$ {\it via} the construction of the Hamiltonian for a $p$-brane
in a YM and antisymmetric tensor background.

In section 2 we shall begin our presentation with an explanation of the method
used. For this purpose we shall consider in detail the special cases of the
string (recovering previous results) and the three-brane (since this is the
simplest illustration of the new features that occur beyond $p=1$).  We then
proceed, in section 3, to a discussion of the general case and in section 4 we
construct the $p$-brane Hamiltonian and find the covariant derivatives.
\bigskip
\noindent{\bf 2. Strings and Three-branes}

We begin with the $p=1$ case, {\it i.e.}\ a closed string moving on
$M\times G$. We first introduce some notation. Let $x^\mu$ and $y^m$ be
local coordinates on $M$ and $G$ respectively and let $A_\mu(x)=A_\mu^aT_a$
and $B_{\mu\nu}(x)$ be background YM and tensor gauge fields on $M$. Let us
denote by $L_a=L_a{}^m(y)\p_m$ the left-invariant vector fields on $G$;
they satisfy $[L_a, L_b]=L_cf^c{}_{ab}$ where $f^c{}_{ab}$ are the
structure constants of $G$. We shall also need a background Riemannian
metric $g_{\mu\nu}$ on $M$ and an invariant Riemannian metric
$g_{mn}=L_m{}^a L_n{}^b d_{ab}$ on $G$, where $d_{ab}= \tr (T_aT_b)$ is a
multiple of the Cartan-Killing inner product on the Lie algebra of $G$. We
also introduce a potential $b_{mn}$ on $G$ chosen to satisfy the relation $
3\p_{[m} b_{np]}= f_{abc}L_m{}^aL_n{}^b L_p{}^c$, where
$f_{abc}=d_{ad}f^d{}_{bc}$ and it is to be understood in what follows that
all lowering and raising of the group indices will be done with $d_{ab}$
and its inverse.

Introducing local worldsheet coordinates $\xi^i=(\tau,\sigma)$ and an
(independent) worldsheet metric $\gamma_{ij}(\xi)$ (with inverse
$\gamma^{ij}$), we can now write the string action as
$$ \eqalign{
S = \int\!\! d^2\xi\, \Bigl[
&-\ft12 \sqrt{-\gamma}\gamma^{ij}\Bigl(\p_i x^\mu\p_j x^\nu g_{\mu\nu}
+(D_iy)^m (D_jy)^n g_{mn}\Bigr)\cr
&\qquad\qquad\quad\quad+ \epsilon^{ij}\left(\ft12 B_{ij}
- k_1L_i^aA_{ja} -\ft12k_1b_{ij} \right)\Bigr]\cr}
\eqno (2.1)
$$
where $B_{ij}, b_{ij}, L_i^a$ and $A_i^a$ are the pullbacks to the
worldsheet of $B_{\mu\nu}$, $b_{mn}$, $L_m^a$ and $A_\mu^a$, respectively,
$\gamma=\det\gamma_{ij}$, and  $(D_i y)^m\equiv \partial_i y^m -\partial_i
x^\mu A_\mu^a L_a^m$ is a YM-covariant derivative. The coefficient
$k_1$ is a normalisation constant; in order that
$\exp(-\ft12ik_1\int\epsilon^{ij}b_{ij})$ be well-defined in the quantum
theory, the coefficient $k_1$ is restricted to be an integer
multiple of some numerical factor, whose precise value will not
concern us here. The last two terms in (2.1) are not
YM invariant by themselves, but their variation can be cancelled, {\sl up
to a total derivative}, by an anomalous variation of $B_{\mu\nu}$.
Specifically, under YM and antisymmetric tensor gauge transformations the
fields transform as $$
\eqalign{ \delta x^\mu =&\ 0\ ,\cr \delta y^m =&\ \epsilon^a (x)
L_a{}^m(y)\ ,\cr \delta A_\mu^a =&\ \partial_\mu\epsilon^a + f^a{}_{bc}
A_\mu^b\epsilon^c =(D_\mu\epsilon)^a\ ,\cr
\delta B_{\mu\nu}=&-2k_1A_{[\mu}^a\p_{\nu]}\epsilon_a
+2\p_{[\mu}\Lambda_{\nu]}\ . \cr} \eqno(2.2)
$$
The variation of the action (2.1) under these
transformations is
$$
\eqalign{
\delta S &=\int\!\! d^2\xi\, \varepsilon^{ij}\partial_i\big[k_1
 \epsilon^a\partial_jy^m L_m{}^b (d_{ab}-b_{ab}) +\partial_j x^\mu
\Lambda_\mu\big]\cr
&=\int\!\! d\tau\partial_\tau\oint\!\!
 d\sigma\big[k_1\epsilon^a{\p_\sigma y}{}^mL_m{}^b(d_{ab}-b_{ab})
+\p_\sigma x{}^\mu\Lambda_\mu \big]\ ,\cr} \eqno (2.3)
$$
where the second line follows from the fact that a closed string has no
boundary. We remark that the inclusion of the WZ term in the action,
with the consequent complications, is not obligatory from a worldsheet
point of view, but it {\sl is} required for the background gauge field
transformations needed for invariance of the action to coincide with
those known from the $D$-dimensional supergravity/YM theory, at least for
the particular cases (D=10, p=1,5) of most interest to us here.

Our aim now is to determine the transformation properties of the string
wave-functional. Let us suppose the world-sheet to be a two-manifold with an
$S^1$ boundary component representing a closed string at a given time, and
consider the string wave-functional
$$
\Psi[x,y;A,B] = \int^{x,y} [dX][dY] e^{iS[X,Y;A,B]}\ ,
\eqno (2.4)
$$
where the arguments $\big(x^\mu(\sigma),y^m(\sigma)\big)$ are the boundary
values of the integration variables\break
$\big(X^\mu(\tau,\sigma),Y^m(\tau,\sigma)\big)$. In the spirit of the `no
boundary' proposal [14] of quantum cosmology, we can avoid having to deal
with boundary conditions at an earlier time by supposing that there is no
such boundary. The consistency of this viewpoint requires that $iS$ be
replaced by minus the Euclidean action, obtained by analytic continuation
of the worldsheet metric from Lorentzian to Euclidean signature, but the
relevant terms in the action are metric-independent and therefore
unaffected by the difference of signature. Now, assuming an invariant
path-integral measure and keeping only first-order variations,
$$
\eqalign{ \Psi[x,y+\delta y;A+\delta A,B+\delta B]
&= \int^{x,y+\delta y}[dX][dY] e^{iS[X,Y;A+\delta A,B+\delta B]}\cr
&=\int^{x,y}[dX][dY] e^{iS[X,Y+\delta Y;A+\delta A,B+\delta B]}\cr
&=\int^{x,y}[dX][dY] e^{i(S+\delta S)}\cr
&=(1+i\delta S)\Psi[x,y;A,B]\ ,\cr}
\eqno (2.5)
$$
where the last line follows from the asssumption that $\delta S$ is a surface
term and therefore independent of the integration variables. The
assumption of the invariance of the measure is justified for the bosonic string
considered here. If we were dealing with the formulation in which the YM fields
couple to the worldsheet {\it via} heterotic fermions then the surface term
would arise not from the variation of the classical action but from the
non-invariance of the measure; however, the final result would be the same.
Expanding the left-hand-side of (2.5) to first order in variations we deduce
that $\Psi$ satisfies the functional differential equation
$$
\eqalign{
\bigg\{\int\!\! d^Dx'&\bigg[\delta A_\mu^a(x'){\delta\over\delta A_\mu^a(x')} +
\delta B_{\mu\nu}(x'){\delta\over\delta B_{\mu\nu}(x')}\bigg] \cr
&+ \oint\! d\sigma \delta
y^m(\sigma){\delta \over\delta y^m(\sigma)}\  -i\ \delta
 S(x,y)\bigg\}\Psi[x,y;A,B]=0\ .\cr}
\eqno (2.6)
$$
where $D$ is the dimension of $M$. The functional derivatives here,
and henceforth, are defined to be densities; for example,
$\delta y^n(\sigma')/\delta y^m(\sigma)=\delta^n_m\delta(\sigma-\sigma')$
where the delta function is a density. Substituting the particular variations
of (2.2) into (2.6) and taking into account the independence of $\epsilon^a$
and
$\Lambda$, we find that
$$
G_\epsilon\Psi(x,y;A,B)=0\qquad
 G_\Lambda\Psi(x,y;A,B)= 0 \eqno (2.7)
$$
where
$$
\eqalignno{ G_\epsilon &=\int \!\! d^Dx'
\bigg\{ \big(D_\mu\epsilon(x')\big)^a{\delta\over\delta A^a_\mu(x')}
-2k_1\big(\partial_\nu\epsilon(x')\big)_a A_\mu^a(x'){\delta\over\delta
B_{\mu\nu}(x')}\bigg\}& \cr
& \qquad \qquad +\oint\!\! d\sigma
\epsilon^a\big(x(\sigma)\big)D_a(\sigma) &(2.8a) \cr G_\Lambda
&=\int\!\! d^Dx'\; 2\partial_\mu\Lambda_\nu(x'){\delta\over\delta
B_{\mu\nu}(x')}\ - i\oint\!\! d\sigma\;
\p_\sigma x{}^\mu(\sigma)\Lambda_\mu\big(x(\sigma)\big)
 &(2.8b)\cr}
$$
with $D_a(\sigma)$ given by
$$
D_a(\sigma)=L_a{}^m{\delta\over\delta y^m(\sigma)}
-ik_1\p_\sigma y{}^mL_m{}^b(d_{ab}-b_{ab})\ .
\eqno (2.9)
$$
Eqs.\ (2.7) state that the wavefunctional (2.4) is invariant under the
transformations (2.2) up to a phase. Although this result was found for a
particular wavefunctional, its general validity is clearly required for
physical quantities to be gauge-independent.

The computation of the algebra of the generators of eqs.\ (2.8) is greatly
simplified by the
fact that the operators $D_a(\sigma)$ are background-field independent. As
shown in [7], these operators generate an affine Ka{\v c}-Moody algebra with
a central extension.  Using this result, we find that all commutators of the
complete generators vanish except for (1.2) with $[\epsilon_1,\epsilon_2]\equiv
f^a{}_{bc}\epsilon_1^b\epsilon_2^c T_a$ and $\Lambda_\mu=
k_1\epsilon_2^a{\buildrel \leftrightarrow\over\partial_\mu }\epsilon_{1a}$.

Before proceeding to the general odd-$p$ case, we shall discuss the closed
three-brane as this provides the simplest illustration of the complications
that
arise beyond $p=1$. The action of [4] describing the coupling of a three-brane
to YM fields exists only for those groups for which there is a third order
symmetric invariant tensor $d_{abc}$ (which satisfies the identity
$f^d{}_{e(a}d_{bc)d}=0$); a simple example would be $SU(3)$. Introducing
worldvolume coordinates $\xi=(\tau;\sigma^r, r=1,2,3)$, we can write the
action as
$$ \eqalign{ S =
\int\!\! d^4\xi\Bigg\{&-\ft12
\sqrt{-\gamma}\gamma^{ij}\big[ \p_i x^\mu\p_j x^\nu g_{\mu\nu}
+d_{ab}(D_iy)^a (D_jy)^b\big] +\sqrt{-\gamma}\cr
&-{k_3\over 8}\varepsilon^{ijkl}d_{abc}\bigg[f^a{}_{de}\Bigl( L_i{}^bA_j^c
A_k^dA_l^e - L_i{}^c L_j{}^dA_k^bA_l^e - L_i{}^cL_j{}^dL_k{}^e
A_l^b\Bigr) +4L_i{}^a\partial_jA_k^bA_l^c\bigg]\cr
&+{1\over 24}\varepsilon^{ijkl}\Big(B_{ijkl}-k_3b_{ijkl}\Big)\Bigg\}\
,\cr} \eqno (2.15)
$$
where $k_3$ is a normalisation constant, and $B_{ijkl}$ and $b_{ijkl}$ are
the pull-backs of $B_{\mu\nu\rho\sigma}$ and $b_{mnpq}$ respectively, the
latter being chosen to satisfy
$$
5\partial_{[m}b_{npqr]}= -{3\over 2} d_{st[m}f^s{}_{np}f^t{}_{qr]}\ .
\eqno (2.16)
$$
The YM and antisymmetric tensor gauge transformations are now
$$
\eqalign{
\delta y^m &= \epsilon^a (x) L_a{}^m(y)\cr
\delta A_\mu^a &=(D_\mu\epsilon)^a\cr
\delta B_{\mu\nu\rho\sigma}&=-12k_3d_{abc}\Big( A_{[\mu}^a\partial_\nu
A_\rho^b +{1\over4}f^a{}_{de}A_{[\mu}^d A_\nu^e
A_\rho^b\Big)\partial_{\sigma]}\epsilon^c
+4\partial_{[\mu}\Lambda_{\nu\rho\sigma]}\ .\cr} \eqno (2.17)
$$
The variation of the action under these transformations is
$$
\eqalign{
\delta S &=\int\!\!
d^4\xi\,\varepsilon^{ijkl}\partial_i\bigg\{-{1\over 6}k_3\epsilon^a\bigg[
L_j{}^b L_k{}^c L_l{}^d(b_{abcd} +{3\over 4} d_{eab} f^e{}_{cd})
+3d_{abc}\partial_l(L_j{}^b A_k^c)\bigg]\cr &\hskip 3cm +
{1\over6}\partial_j x^\mu\partial_k x^\nu \partial_l x^\rho
\Lambda_{\mu\nu\rho}\bigg\}\ .\cr} \eqno (2.18)
$$
Since the three-brane has no boundary, this surface term can be written as
$$
\eqalign{
\delta S =\int\!\! d\tau\partial_\tau\oint d^3\sigma\,
 \varepsilon^{rst}\bigg\{&-{1\over 6}k_3\epsilon^a
\big[L_r{}^bL_s{}^cL_t{}^d(b_{abcd}+{3\over 4}
d_{eab}f^e{}_{cd}) + 3d_{abc}\partial_t(L_r{}^bA_s^c)\big]\cr
& +{1\over 6}\partial_r x^\mu\partial_s x^\nu\partial_t x^\rho
\Lambda_{\mu\nu\rho}\bigg\}\ .\cr}
\eqno (2.19)
$$
This result leads, by the same reasoning as before, to the
generators
$$
\eqalignno{
G_\epsilon &=\int \!\! d^Dx'\bigg\{
\big(D_\mu\epsilon(x')\big)^a{\delta\over\delta A^a_\mu(x')}
-12k_3d_{abc} \big(
A_{\mu}^a\partial_\nu A_\rho^b +{1\over4}f^a{}_{de}A_{\mu}^d A_\nu^e
A_\rho^b\big) \big(\partial_\sigma\epsilon^c(x')\big) {\delta\over\delta
B_{\mu\nu\rho\sigma} (x')} \bigg\}\cr
&\qquad +\oint\! d^3\sigma\; \epsilon^a\big(x(\bs)\big)D_a(\bs)
&(2.20a)\cr G_\Lambda &=\int\!\! d^Dx'\,
4\partial_\mu\Lambda_{\nu\rho\sigma}(x'){\delta\over\delta
B_{\mu\nu\rho\sigma}(x')} -i\oint\! d^3\sigma\;
\varepsilon^{rst}\partial_r x^\mu \partial_s x^\nu \partial_t x^\rho
\Lambda_{\mu\nu\rho}\big(x(\bs)\big) &(2.20b)\cr}
$$
where $\bs=\{\sigma^r\}$, and
$$
D_a(\bs)= L_a{}^m{\delta\over\delta y^m(\bs)}
+{ik_3\over 6}\varepsilon^{rst}
\big[ L_r{}^bL_s{}^cL_t{}^d(b_{abcd}+{3\over 4} d_{eab}f^e{}_{cd})
+3d_{abc}\partial_t(L_r{}^bA_s^c)\big]  \eqno (2.21)
$$
Observe that, in contradistinction to the string case, the operators $D_a(\bs)$
{\sl depend on the background fields} so that their algebra differs from the
algebra of the complete generators, and therefore has no obvious significance.
However, the background-field dependence of $D_a(\bs)$ is such that the algebra
of the complete generators, $G_\epsilon$ and $G_\Lambda$, is the same as (1.2)
with $[\epsilon_1,\epsilon_2]\equiv f^a{}_{bc}\epsilon_1^b\epsilon_2^c T_a$ and
$\Lambda_{\mu\nu\rho}=k_3d_{abc}\partial_{[\mu}\epsilon_1^a\partial_\nu\epsilon_2^b
A_{\rho]}^c$. For a related discussion of the algebra of gauge
transformations in the context of field-theoretic gauged Wess-Zumino
terms, see Ref.\ [11].

\bigskip
\noindent{\bf 3. The general odd-$p$ case}

We now turn to the case of a closed $p$--brane propagating in $M\times G$ for
$p$ odd but otherwise arbitrary. The background fields on $M$ are the metric
$g_{\mu\nu}(x)$, the antisymmetric tensor field $B_{\mu_1\dots\mu_{p+1}}(x)$,
and the YM field $A_\mu =A_\mu^a(x)T_a$ valued in some representation of the
Lie algebra of $G$.  The background fields on
$G$ are now the left-invariant metric $g_{mn}(y)$ and the potential
$b_{m_1\dots m_{p+1}}(y)$ satisfying
$$
\p_{[m_1}b_{m_2\ldots m_{p+2}]}=
-c_p(p+1)!{\rm tr}L_{[m_1}\ldots L_{m_{p+2}]}\ ,
\eqno(3.1)
$$
where $L_m=L_m{}^a T_a$, and $c_p$ is the constant
$c_p=(-1)^{{p+1}\over 2}\left({{p+3}\over2}\right)
\Gamma({{p+3}\over2})^2/\Gamma(p+3)$.

We may now write the action (for unit $p$-volume tension) as [4]
$$
\eqalign{
S = \int\!\! d^{p+1}\xi\, &\bigg\{ -\ft12\sqrt{-\gamma}\gamma^{ij}
\left(\p_i x^\mu\p_j x^\nu g_{\mu\nu}
+D_iy^m D_jy^n g_{mn}\right)+{(p-1)\over2}\sqrt{-\gamma}\cr
&\qquad\qquad
+{1\over (p+1)!}\varepsilon^{i_1\dots i_{p+1}}\big[B_{i_1\dots i_{p+1}}
+k_pC_{i_1\dots i_{p+1}}-k_pb_{i_1\dots i_{p+1}}\big]\bigg\}\ ,\cr}
\eqno(3.2)
$$
where $k_p$ is a normalisation constant, $B_{i_1\dots i_{p+1}}$ and
$b_{i_1\dots i_{p+1}}$ are the pullbacks to the worldvolume of the
corresponding antisymmetric tensors on $M$ and $G$, and $C_{i_1\dots
i_{p+1}}$ are the components of the pullback of a $(p+1)$-form $C_{p+1}$
on $M\times G$ that is constructed as follows. Let $A_t=tA+(1-t)L$ and
$F_t=dA_t+A_t^2$, where $A=dx^\mu A_\mu$ and $L=dy^m L_m$. Defining the
operator $$
\ell_t= dt (A^a-L^a){\partial\over\partial F_t^a}\ ,
\eqno(3.3)
$$
we have
$$
C_{p+1}(A,F,L)=\int_0^1\! \ell_t\ \omega_{p+2}^0(A_t,F_t)\ ,
\eqno(3.4)
$$
where $\omega_{p+2}^0$ is the Chern-Simons form defined by the relation
$$
d\omega_{p+2}^0={\rm tr}F^{{p+3}\over2}.
\eqno(3.5)
$$
The cases of most interest are $p=1,3,5$, for which
$$
\eqalign{
\omega_3^0(A,F) &= {\rm tr} \bigg(FA -{1\over 3}A^3\bigg)\cr
\omega_5^0(A,F) &= {\rm tr} \bigg(F^2A -\h F A^3+{1\over 10}A^5\bigg)\cr
\omega_7^0(A,F) &= {\rm tr } \bigg(F^3A -{2\over 5}F^2A^3
-{1\over 5}FA FA^2 +{1\over 5}FA^5 -{1\over 35}A^7\bigg)\ ,\cr}
\eqno (3.6)
$$
and substituting these expressions into (3.4) we find that
$$
\eqalign{
C_2=&\,\tr(AL)\cr
C_4=&{1\over4}\tr\left[2(FA+AF-A^3)L+ ALAL-2AL^3\right]\cr
C_6=&{1\over30}\tr\bigl[
(10F^2A+10FAF+10AF^2-8FA^3-8A^3F-4AFA^2-4A^2FA+6A^5)L\cr
&+2F(A^2L^2-L^2A^2+3ALAL-3LALA)-6A^3LAL\cr
&+3F(LAL^2-L^2AL+2L^3A-2AL^3)+6A^3L^3\cr
&-3L^2A^2LA+3A^2L^2AL+2ALALAL+6L^3ALA+6AL^5\bigr]\ .\cr}
\eqno (3.7)
$$
The three-brane action of (2.15) agrees with (3.2) if the identification
$d_{abc}=\tr\big(T_a\{T_b,T_c\}\big)$ is made. The YM gauge variation of the
Chern--Simons forms defines the $(p+1)$-form $\omega^1_{p+1}$:
$$
\delta_\epsilon\omega^0_{p+2}(A,F)=d\omega^1_{p+1}(A,F,\epsilon)\ .
\eqno (3.8)
$$
As explained in [15], $\omega^1_{p+1}$ can be written in the form
$$
\omega^1_{p+1}(A,F,\epsilon)=\tr\,d\epsilon\,\phi_p(A,F)
\eqno (3.9)
$$
where the $p$-form $\phi_p=\phi_p^a T_a$ is a Lie algebra-valued polynomial
in $A$ and $F$, given, for $p=1,3,5$, by
$$
\eqalign{
\phi_1=&\ -A\cr
\phi_3=&\ -\ft12(FA+AF-A^3)\cr
\phi_5=&\ -\ft13\left[(F^2A+FAF+AF^2)-\ft45(A^3F+FA^3)
-\ft25(A^2FA+AFA^2)+\ft35 A^5\right]\cr}
\eqno (3.10)
$$
The components of the form $\omega^1_{p+1}$ appear in the YM transformation of
the antisymmetric tensor field; the full YM and antisymmetric tensor gauge
transformations are
$$
\eqalign{
\delta x^\mu=&0\cr
\delta y^m=&\,\epsilon^a(x)L^m_a(y)\cr
\delta A_\mu^a=&\,\p_\mu\epsilon^a+f^a{}_{bc}A_\mu^b \epsilon^c\cr
\delta
B_{\mu_1\ldots\mu_{p+1}}=&\,-k_p\omega^1_{\mu_1\ldots\mu_{p+1}}(A,F,\epsilon)
+(p+1)\partial_{[\mu_1} \Lambda_{\mu_2\ldots\mu_{p+1}]} \cr}
\eqno(3.11)
$$

The variation of $C_{p+1}$ is determined as follows. Note first that $\ell_t$
commutes with $\delta_\epsilon$ since $A-L$ and $F$ both transform
homogeneously. Then, from the definition (3.4) of $C_{p+1}$, and (3.8) it
follows that
$$
\delta_\epsilon C_{p+1} = \int_0^1\! \ell_t\
d\omega_{p+1}^1(A_t,F_t,\epsilon)\ . \eqno (3.12)
$$
Now apply the homotopy formula [15]
$$
d_t \equiv dt {d\over dt} = \ell_td-d\ell_t\ ;
\eqno (3.13)
$$
this leads to the formula
$$
\delta_\epsilon C_{p+1} =
\omega_{p+1}^1(A,F,\epsilon)-\omega_{p+1}^1(L,\epsilon) + d\int_0^1\! \ell_t
\tr\,d\epsilon \phi_p(A_t,F_t) \eqno (3.14)
$$
where $\omega^1_{p+1}(L,\epsilon)=\omega^1_{p+1}(L,0,\epsilon)$, and we have
 rewritten
the last term using (3.9). Finally, defining a Lie algebra-valued $(p-1)$-form
$\chi_{p-1}=\chi_{p-1}^aT_a$ by
$$
\chi_{p-1} = \int_0^1\! \ell_t\ \phi_p(A_t,F_t)\ ,
\eqno (3.15)
$$
we obtain
$$
\delta_\epsilon k_pC_{p+1}
=k_p\Big(\omega^1_{p+1}(A,F,\epsilon)-\omega^1_{p+1}(L,\epsilon) +
d\big(\tr\,[\epsilon d\chi_{p-1}]\big)\Big) \ .\eqno (3.16)
$$
The variation of the last term in (3.2) is given by
$$
\eqalign{
& -{k_p\over p!}\epsilon^{i_1\dots i_{p+1}}\partial_{i_1}
\biggl \{ \epsilon^aL_a^{m_1}\partial_{i_2}y^{m_2}\dots
\partial_{i_{p+1}}y^{m_{p+1}}b_{m_1m_2\dots m_{p+1}}\biggr \} \cr
& -{(p+2)k_p\over (p+1)!}\epsilon^{i_1\dots
i_{p+1}}\epsilon^aL_a^n\partial_{i_1} y^{m_1}\dots
\partial_{i_{p+1}}y^{m_{p+1}}\partial_{[n}b_{m_1\dots m_{p+1}]}\ .\cr}
\eqno (3.17)
$$
The first term here is a surface term; using (3.1), the second term is
seen to be equal to
$$
k_pc_p(p+2)\tr\; \epsilon L^{p+1} =k_p\tr\; \epsilon d\phi_p(L) =
k_p\Big(\omega^1_{p+1}(L,\epsilon)-d(\tr\; \epsilon\phi_p(L))\Big)\ ,
\eqno(3.18)
$$
where we have used the fact that $\phi_p(L)\equiv \phi_p(L,0)=c_p (p+2) L^p$
and then (3.9).  One can now easily prove the invariance of the action {\sl up
to a surface term}. The terms in the first line of (3.2) are manifestly
invariant. The variation of $B$ cancels the first term on the r.h.s.\ of
(3.16).
The first term on the r.h.s.\ of (3.18) cancels the second term on the r.h.s.\
of (3.16).

All that remains are the first term in (3.17), the third term on the r.h.s.\ of
(3.16) and the second term on the r.h.s.\ of (3.18). Also taking into account
the surface terms coming from the tensor gauge transformations, we find
$$
\eqalign{\delta S &=
\int\!\! d\tau\partial_\tau\oint\!\! d^p\sigma
{1\over p!}\epsilon^{r_1\dots r_p}\Bigg\{\Lambda_{r_1\dots r_p}\cr
&-k_p\epsilon^a\Big[ L_a{}^m b_{mr_1\dots r_p}
+d_{ab}\bigl( \phi_p^b(L)\bigr)_{r_1\dots r_p}
-pd_{ab}\partial_{[r_1}\bigl(\chi^b_{p-1}\bigr)_{r_2\dots r_p]}\Big]\Bigg\}\
 ,\cr}
\eqno(3.19)
$$
where we have decomposed the $i$-index into a time and a space part according
to
$i=(0,r_1\dots r_p)$. In this formula, appropriate pull-backs with $\partial_r
x^\mu$ and $\partial_r y^m$ are understood. (For comparison with the string and
three-brane formulae of section 3 we note here that $\chi_0^a=0$,
$\chi_2^a=-{1\over 2}d^a{}_{bc}A^bL^c$; also note
$\chi_4^a=d^a{}_{bcd}L^bA^c[F^d-\ft1{10}f^d{}_{fg}(3A^fA^g+3L^fL^g-2L^fA^g)]$,
where $d_{abcd}=\tr(T_{(a}T_bT_cT_{d)})$).

Following the reasoning of the string and three-brane examples we deduce from
(3.19) that the generators acting on the $p$--brane wavefunctional are given by
$$
\eqalignno{
G_\epsilon =\int\!\! d^Dx'\, \bigg\{\big(D_\mu\epsilon^a(x')\big)
&{\delta\over\delta A_\mu^a(x')}
-(p+1)k_p\big(\partial_{\mu_1}\epsilon^a(x')\big)d_{ab}
 (\phi_p^b)_{\mu_2\ldots\mu_{p+1}}(A,F)
{\delta\over\delta B_{\mu_1\ldots\mu_{p+1}}(x')}\bigg\}\cr
&\qquad +\oint\!d^p\sigma \;\epsilon^a D_a(\bs) &(3.20a)\cr}
$$
$$\eqalign{
G_\Lambda =\int \!\! d^Dx'&\, \bigg\{
(p+1)\partial_{\mu_1}\Lambda_{\mu_2\ldots\mu_{p+1}}(x')
{\delta\over\delta B_{\mu_1\ldots\mu_{p+1}}(x')}\bigg\}\cr
&-i\oint\! d^p\sigma\;
\varepsilon^{r_1\ldots r_p} \p_{r_1}x^{\mu_1}\cdots\p_{r_p}x^{\mu_p}
\Lambda_{\mu_1\ldots\mu_p}\ ,\cr}\eqno(3.20b)
$$
where $D_a(\bs)$ is given by
$$
D_a(\bs)= L_a{}^m \Bigg[{\delta\over\delta y^m(\bs)}
+i {k_p\over p!}\varepsilon^{r_1\ldots r_p} \Bigg\{
b_{m r_1\ldots r_p}+
L_{mb} \bigg[
\bigl(\phi_p^b(L)\bigr)_{r_1\dots r_p}-
p\partial_{[r_1}\bigl(\chi_{p-1}^b\bigr)_{r_2\dots r_p]}\bigg]\Bigg\}
\Bigg]
\eqno (3.21)
$$

We refer to [16] for a computation of the algebra of these generators,
which is that given in (1.2). We note here that once it is known that they
form a {\sl closed} algebra, (1.2) follows from the transformations (3.11)
because $G_\epsilon$ and $G_\Lambda$ reproduce these transformations when
acting on the individual fields.
\bigskip
\noindent{\bf 4. Covariant Derivatives in $p$--Loop Space}

The coupling to the antisymmetric tensor $B_{\mu\nu}$ and the WZ term in
the action (3.2) are linear in time derivatives and can therefore be expressed
 as
$$
\int\! d\tau\oint\! d^p\sigma\big[\dot x^\mu {\bf B}_\mu -k_p\dot y^m {\bf
b}_m\big]\ , \eqno (4.1)
$$
where the overdot indicates differentiation with respect to $\tau$.
Using the chain rule for $C(A,F,L)$ we have that
$$
\eqalign{{\bf B}_\mu &={1\over p!}\varepsilon^{r_1\dots
r_p}\partial_{r_1}x^{\nu_1}\cdots \partial_{r_p} x^{\nu_p}
 B_{\mu\nu_1\dots\nu_p}
+k_pA_\mu^a{\partial C\over \partial A_0^a} +k_pF_{\mu\nu}^a\partial_r
x^\nu{\p C\over \partial F_{0r}^a}\cr
{\bf b}_m &={1\over p!}\varepsilon^{r_1\dots r_p}\partial_{r_1}y^{n_1}
\cdots\partial_{r_p}y^{n_p}b_{m n_1\dots n_p}
- L_m{}^a{\partial C \over \partial L_0{}^a}\ .\cr}
\eqno (4.2)
$$
Introducing the momenta $p_\mu(\tau,\bs)$ and $p_m(\tau,\bs)$, conjugate to the
worldvolume fields $x^\mu$ and $y^m$ respectively, and the Lagrange multipliers
$\ell(\tau,\bs)$ and $s^r(\tau,\bs)$ for, respectively, the time and space
reparametrization constraints, we can rewrite the action (3.2) in the
equivalent
`first-order' form
$$
S=\int\!d\tau\oint\!d^p\sigma\big[ \dot x^\mu p_\mu +\dot y^m p_m -\ell\; {\cal
 H}_0
-s^r\; {\cal H}_r\big]\ .
\eqno (4.3)
$$
The constraint functions are
$$
\eqalign{ {\cal H}_0 &= g^{\mu\nu}{\cal P}_\mu {\cal P}_\nu +d^{ab}{\cal
 P}_a{\cal P}_b
+\det \big( \p_r x^\mu\p_s x^\nu g_{\mu\nu} + (D_r y)^m(D_s y)^n g_{mn}\big)\cr
{\cal H}_r &= \p_r x^\mu {\cal P}_\mu +(D_r y)^mL_m{}^a {\cal P}_a\cr}
\eqno (4.4)
$$
where
$$
\eqalign{
{\cal P}_\mu &= p_\mu -{\bf B}_\mu + A_\mu^a {\cal P}_a\cr
{\cal P}_a &= L_a{}^m(p_m +k_p{\bf b}_m)\cr}
\eqno (4.5)
$$
The (classical) equivalence of (4.3) to (3.2) can be proved by eliminating all
auxiliary variables from both actions.

Observe that the action (4.3) can be rewritten as
$$
\eqalign{
S=\int\!d\tau\oint\!d^p\sigma\big[ &\dot x^\mu {\cal P}_\mu
 +(D_0y)^mL_m{}^a{\cal P}_a
-\ell\; {\cal H}_0 -s^r\; {\cal H}_r\big]\cr
&+\int\!d\tau\oint\!d^p\sigma\big[\dot x^\mu{\bf B}_\mu -k_p\dot y^m{\bf
b}_m\big]\
 .\cr}
\eqno (4.6)
$$
The second integral is just (4.1) which was shown in section 3
to be invariant, up to a surface term, under the gauge transformations (3.11).
The first integral is invariant if these transformations are supplemented by
transformations of $p_\mu$ and $p_m$ chosen such that the functions ${\cal
P}_\mu$ and ${\cal P}_a$ are covariant, {\it i.e.}\ such that $$
\delta {\cal P}_\mu =0 \qquad \qquad \delta {\cal P}_a = -{\cal P}_c f^c{}_{ab}
\epsilon^b \ .
\eqno (4.7)
$$
The corresponding statement in the quantum theory, obtained by the
replacements
$$
p_\mu(\bs)\rightarrow \hat p_\mu(\bs)=-i{\delta\over\delta x^\mu(\bs)} \qquad
p_m(\bs)\rightarrow \hat p_m(\bs)=-i{\delta\over\delta y^m(\bs)}\ ,
\eqno (4.8)
$$
is that the operators ${\cal D}_\mu\equiv i\hat{\cal P}_\mu$ and ${\cal
D}_a\equiv i\hat{\cal P}_a$ obey (1.5).

Of chief importance is the derivative ${\cal D}_\mu$. Using  (4.2) and (4.5) we
find that
$$
\eqalign{
{\cal D}_\mu &= {\delta\over\delta x^\mu(\bs)} -{i\over
p!}\varepsilon^{r_1\dots r_p}\p_{r_1} x^{\nu_1}\cdots\p_{r_p}x^{\nu_p}
B_{\mu\nu_1\dots \nu_p} -ik_pF_{\mu\nu}^a\p_r x^\nu{\p C\over \p
F_{0r}^a}\cr
&+A_\mu^a L_a{}^m\bigg[
{\delta\over \delta y^m(\bs)} +{ik_p\over p!}\varepsilon^{r_1\dots
r_p}\p_{r_1}y^{n_1}\cdots\p_{r_p}y^{n_p} b_{mn_1\dots n_p}
-k_pL_m{}^b\bigg({\p C\over\p L_0{}^b} +{\p C\over\p A_0^b}\bigg)\bigg]\
.\cr} \eqno (4.9) $$
It can be shown [16] that
$$
{\p C\over\p A_0^a}+{\p C\over\p L_0{}^a}= d_{ab}\bigg(
\phi^b_p(A,F)-\phi^b_p(L)+d\chi^b_{p-1}\bigg)\ .
\eqno (4.10)
$$
from which it follows that
$$
{\cal D}_\mu = {\delta\over\delta x^\mu(\bs)} -i{\cal C}_\mu +A_\mu^a
D_a(\bs) \eqno (4.11)
$$
where $D_a(\bs)$ is the $p$--loop space differential operator given previously
in (3.21) and
$$
{\cal C}_\mu = {1\over p!}\varepsilon^{r_1\dots r_p}\p_{r_1}
x^{\nu_1}\cdots\p_{r_p}x^{\nu_p} B_{\mu\nu_1\dots \nu_p} +k_pA_\mu^a
\phi_p^b(A,F) d_{ab} + k_pF_{\mu\nu}^a\p_r x^\nu{\p C\over \p F_{0r}^a}\ ,
\eqno (4.12)
$$
which we identify as the $U(1)$ gauge potential. For $p=1$ this reproduces the
results of [7,8].

\bigskip\bigskip \centerline{\bf Acknowledgements}
\noindent
E.B. thanks DAMTP and Texas A\&M University, R.P. thanks Texas A\&M
University, E.S. thanks the University of Groningen and the International
Center for Theoretical Physics in Trieste, K.S.S. thanks the University
of Groningen and SISSA and P.K.T. thanks the University of Groningen for
hospitality.  The work of E.B. has been made possible by a fellowship of
the Royal Netherlands Academy of Arts and Sciences (KNAW).
\bigskip\bigskip \goodbreak
\centerline{\bf References}
\bigskip
\frenchspacing

\item {[1]}
E. Bergshoeff, B. de Wit, M. de Roo and P. van Nieuwenhuizen, {\sl Nucl.
Phys.} {\bf B195} (1982) 97;
G.F. Chapline and N.S. Manton, {\sl Phys. Lett.} {\bf 120B} (1983) 105.

\item {[2]}
S.J. Gates and H. Nishino, {\sl Phys. Lett.} {\bf 157B} (1985) 157;
A. Salam and E. Sezgin, {\sl Phys. Scripta} {\bf 32} (1985) 283.

\item {[3]}
R. Nepomechie and P.G.O. Freund, {\sl Nucl. Phys.} {\bf B199} (1982) 482;
P.A. Marchetti and R. Percacci, {\sl Lett. Math. Phys.} {\bf 6} (1982) 405;
H. Zainuddin, {\sl J. Math. Phys.} {\bf 31} (1990) 2225;
E. Bergshoeff, P.S. Howe, C.N. Pope, E. Sezgin and E. Sokatchev, {\sl Nucl.
Phys.} {\bf B354} (1991) 113;
J.A. de Azc{\' a}rraga, J.M. Izquierdo and P.K. Townsend, {\sl Phys. Lett.}
{\bf 267B} (1991) 366.

\item {[4]}
J.A. Dixon, M.J. Duff and E. Sezgin, {\sl Phys. Lett.} {\bf 279B} (1992)
265.

\item {[5]}
E. Witten, {\sl Nucl. Phys.} {\bf B 223} (1983) 422;
A. Manohar and G. Moore, {\sl Nucl. Phys.} {\bf B243} (1984) 55;
\"O. Kaymakcalan, S. Rajeev and J. Schechter, {\sl Phys. Rev.} {\bf D 30}
594 (1984); N.K. Pak and P. Rossi {\sl Nucl. Phys.} {\bf B250} (1985) 279;
C.M. Hull and B. Spence, {\sl Nucl. Phys.} {\bf B353} (1991) 379.

\item {[6]}
N.K. Nielsen, {\sl Nucl. Phys.} {\bf B167} (1980) 249;
M.J. Duff, B.E.W. Nilsson and C.N. Pope, {\sl Phys. Lett.} {\bf 163B} (1985)
343; M.J. Duff, B.E.W. Nilsson, C.N. Pope and N.P. Warner, {\sl Phys. Lett.}
{\bf 171B} (1986) 170; R. Nepomechie, {\sl Phys. Lett.} {\bf 171B} (1986) 195;
{\sl Phys. Rev.} {\bf D33} (1986) 3670;
R. Kallosh, {\sl Phys. Scripta} {\bf T15} (1987) 118;
{\sl Phys. Lett.} {\bf 176B} (1986) 50.

\item {[7]}
E. Bergshoeff, F. Delduc and E. Sokatchev, {\sl Phys. Lett.} {\bf 262B}
(1991) 444.

\item {[8]}
P.S. Howe, {\sl Phys. Lett.} {\bf 273B} (1991) 90.

\item {[9]} J.A. Dixon and M.J. Duff, preprint, CTP-TAMU-45/92.

\item {[10]} J. Mickelsson, {\sl Lett. Math. Phys.} {\bf 7} (1983) 45;
{\sl Comm. Math. Phys.} {\bf 97} (1985) 361;\nl
L. Faddeev, {\sl Phys. Lett.} {\bf 145B} (1984) 81.

\item {[11]} R. Percacci and R. Rajaraman, {\sl Phys. Lett.}
{\bf 201B} (1988) 256; {\sl Int. J. Mod. Phys.} {\bf A4} (1989) 4177; T.
Fujiwara, S. Hosono and S. Kitakado, {\sl Mod. Phys. Lett.} {\bf A 3}
(1988) 1585; T. Fujiwara and S. Kitakado, {\sl Z. Phys.} {\bf C 43}
(1989) 201; T. Inamoto, Phys. Rev. {\bf D 45} (1992) 1276.

\item {[12]}
C.M. Hull and E. Witten, {\sl Phys. Lett.} {\bf 160B} (1985) 398.

\item {[13]} J.A. Dixon, preprint, CTP-TAMU-50/92.

\item {[14]} J. B. Hartle and S.W. Hawking, {\sl Phys. Rev.} {\bf D28}
(1983) 2960.

\item {[15]} J. Ma{\~ n}es, R. Stora and B. Zumino,
{\sl Comm. Math. Phys.} {\bf 102} (1985) 157.

\item {[16]} R. Percacci and E. Sezgin, ``Symmetries of $p$-Branes''
preprint, CTP-TAMU-32/92.
\bye